\documentclass[a4paper,11pt]{article}
\usepackage{pos}

\usepackage{caption}
\usepackage{subcaption}
\usepackage{hyperref}
\usepackage{url}

\title{Lattice studies of Sp(2N) gauge theories using GRID}

\author[a]{Ed Bennett}
\author[b,c]{Peter Boyle}
\author[d]{Luigi Del Debbio}
\author*[e]{Niccol\`o Forzano}
\author[f]{Deog Ki Hong}
\author[f,g,h]{Jong-Wan Lee}
\author[a]{Julian Lenz}
\author[i,l,m]{C.-J. David Lin}
\author[a,n]{Biagio Lucini}
\author[d]{Alessandro Lupo}
\author[e]{Maurizio Piai}
\author[o]{Davide Vadacchino}

\affiliation[a]{Swansea Academy of Advanced Computing, Swansea University (Bay Campus),
Fabian Way, SA1 8EN Swansea, Wales, United Kingdom}
\affiliation[b]{School of Physics and Astronomy, University of Edinburgh, Edinburgh EH9 3FD, United Kingdom}
\affiliation[c]{Physics Department, Brookhaven National Laboratory, Upton, NY 11973, USA}
\affiliation[d]{Higgs Centre for Theoretical Physics, School of Physics and Astronomy, The University of Edinburgh, Peter Guthrie Tait Road, Edinburgh EH9 3FD, UK}
\affiliation[e]{Department of Physics, Faculty of Science and Engineering,
Swansea University, Singleton Park, SA2 8PP, Swansea, Wales, UK}
\affiliation[f]{Department of Physics, Pusan National University, Busan 46241, Korea}
\affiliation[g]{Institute for Extreme Physics, Pusan National University, Busan 46241, Korea}
\affiliation[h]{Particle Theory and Cosmology Group, Center for Theoretical Physics of the Universe, Institute for Basic Science (IBS), Daejeon, 34126, Korea}
\affiliation[i]{Institute of Physics, National Yang Ming Chiao Tung University, 1001 Ta-Hsueh Road, Hsinchu 30010, Taiwan}
\affiliation[l]{Centre for High Energy Physics, Chung-Yuan Christian University, Chung-Li 32023, Taiwan}
\affiliation[m]{Centre for Theoretical and Computational Physics,
National Yang Ming Chiao Tung University, 1001 Ta-Hsueh Road, Hsinchu 30010, Taiwan}
\affiliation[n]{Department of Mathematics, Faculty of Science and Engineering, Swansea University (Bay Campus), Fabian Way, SA1 8EN Swansea, Wales, United Kingdom}
\affiliation[o]{Centre for Mathematical Sciences, University of Plymouth, Plymouth, PL4 8AA, United Kingdom}

\emailAdd{2227764@swansea.ac.uk}

\abstract{Four-dimensional gauge theories based on symplectic Lie groups 
provide elegant realisations of the 
 microscopic origin of several new physics models. Numerical studies pursued on
the lattice  provide  quantitative information necessary for phenomenological applications.
To this purpose, we implemented $Sp(2N)$ gauge theories using Monte Carlo techniques within Grid, a performant framework designed for the numerical study of quantum field theories on the lattice.
We show the first results obtained using this library, focusing on the case-study provided by 
the $Sp(4)$ theory coupled to $N_{as} = 4$ Wilson-Dirac fermions transforming in the 2-index antisymmetric representation. In particular, we discuss preliminary tests of the algorithm and we test some of its main functionalities.}

\FullConference{The 40th International Symposium on Lattice Field Theory (Lattice 2023)\\
July 31st - August 4th, 2023\\
Fermi National Accelerator Laboratory\\}

\begin{document}

\newcommand{\Tr}{{\rm Tr}}
\newcommand{\spn}{$Sp(2N)~$}
\newcommand{\spfour}{$Sp(4)~$}

\makeatletter
\newsavebox{\@brx}
\newcommand{\llangle}[1][]{\savebox{\@brx}{\(\m@th{#1\langle}\)}%
  \mathopen{\copy\@brx\mkern2mu\kern-0.9\wd\@brx\usebox{\@brx}}}
\newcommand{\rrangle}[1][]{\savebox{\@brx}{\(\m@th{#1\rangle}\)}%
  \mathclose{\copy\@brx\mkern2mu\kern-0.9\wd\@brx\usebox{\@brx}}}
\makeatother

\maketitle

\section{Introduction}
\label{intro}
Four-dimensional symplectic gauge theories stand out in the literature for their relevance in the context of new physics models.  For this reason, a first quantitative study of the strongly coupled dynamics based on \spn gauge theories was obtained using lattice field theory methods~\cite{Bennett:2017kga, Bennett:2019jzz,Bennett:2019cxd,Bennett:2022yfa,Bennett:2023wjw,Bennett:2023rsl}. The \spfour theory with $N_f = 2$, and $N_{as} = 3$ is particularly interesting \cite{Bennett:2022yfa}: it gives rise, at low energies, to the effective field theory entering the minimal Composite Higgs model \cite{Barnard:2013zea,Ferretti:2013kya} (see  Refs.~\cite{Panico:2015jxa,Cacciapaglia:2019bqz} and references therein), and also realises top (partial) compositeness \cite{Kaplan:1991dc}. In this work, we make preliminary, somewhat technical, progress to lay the foundation for future large-scale studies for \spn  theories using Grid \cite{Boyle:2015tjk,Boyle:2016lbp}. To this end, we wrote and tested new code \cite{sp2ngrid} that supports the study of \spn theories with matter fields in multiple representations. 
This report is structured as follows. In Sect.~\ref{sec:sp2ntheories}, we define the \spn gauge theories of interest, both in the continuum and on the lattice. Sect.~\ref{sec:preliminaryresults} discusses all the tests we performed on the algorithm, and we test some of its main functionalities. Finally, we draw our conclusions in Sect.~\ref{sec:conclusions}.


\section{Symplectic gauge theories and lattice setup}
\label{sec:sp2ntheories}
In the continuum, we consider \spn field theories ($N > 1$), having as Lagrangian densities
\begin{eqnarray}
    {\cal L}&=& -\frac{1}{2} \hbox{Tr} \ G_{\mu\nu} G^{\mu\nu}
\,+\,\frac{1}{2}\sum_i^{N_{\rm f}}\left(i\overline{Q^{i}}_a \gamma^{\mu}\left(D_{\mu} Q^i\right)^a
\,-\,i\overline{D_{\mu}Q^{i}}_a \gamma^{\mu}Q^{i\,a}\right)\,-\,m^{\rm f}\sum_i^{N_{\rm f}}\overline{Q^i}_a Q^{i\,a}+\nonumber\\
&&
\,+\,\frac{1}{2}\sum_k^{N_{\rm as}}\left(i\overline{\Psi^{k}}_{ab} \gamma^{\mu}\left(D_{\mu} \Psi^k\right)^{ab}
\,-\,i\overline{D_{\mu}\Psi^{k}}_{ab} \gamma^{\mu}\Psi^{k\,ab}\right)\,-\,m^{\rm as} \sum_k^{N_{\rm as}}\overline{\Psi^k}_{ab} \Psi^{k\,ab}\,,
\label{eq:lagrangian}
\end{eqnarray}
where the mass-degenerated Dirac fermions $Q^{ia}$, with $i=1,\cdots,N_{\rm f}$, and $a=1,\cdots,2N$, transform in the fundamental representation, whereas $\Psi_{k\,ab}$, with $k=1,\cdots,N_{\rm as}$, transform in the $2$-index antisymmetric representation.  The covariant derivatives are defined through the transformation properties under the action of an element $U$ of the \spn gauge group---$Q \to U Q, \, \Psi \to U \Psi U^T$. \\
We consider the system discretised on a lattice of size $\tilde{V} / a^4   = T \times L^3$, $a$ being the lattice spacing. The action on the lattice is the sum of two terms, $S \equiv S_g + S_f$, where $S_g$ is the gauge action $S_g\equiv\beta \sum_x \sum_{\mu<\nu} \left(1-\frac{1}{2N} {\rm Re}\, {\hbox{Tr}}\, 
{\cal P}_{\mu\nu}(x)
\right)$
where ${\cal P}_{\mu\nu}(x)\equiv U_\mu (x) U_\nu (x+\hat{\mu}) U_\mu^\dagger(x+\hat{\nu}) U_\nu^\dagger(x)$ is
the \emph{elementary plaquette}, $U_{\mu}(x) \in Sp(2N)$ is the
\emph{link variable}, and
$\beta\equiv 4N/g_0^2$ is the \emph{inverse bare gauge coupling}. The fermion action $S_f$ is $S_f \equiv a^4 \sum_{j=1}^{N_{\rm f}}\sum_x \overline{Q}^j(x) D^{\mathrm{(f)}}_m Q^j(x)+
a^4 \sum_{j=1}^{N_{\rm as}}\sum_x \overline{\Psi}^j(x) D^{\mathrm{(as)}}_m \Psi^j(x)$, where the covariant derivatives $D^{(f)}_m$ and $D^{(as)}_m$ are built using the links in the fundamental and 2-index antisymmetric representations, respectively. We shall indicate with $m_0^{\mathrm{f}}$ and $m_0^{\mathrm{as}}$ are the bare masses of the fermions in the fundamental and $2$-index antisymmetric representation, 


\begin{figure}
     \centering
     \begin{subfigure}[b]{0.45\textwidth}
         \centering
         \includegraphics[width=0.9\textwidth]{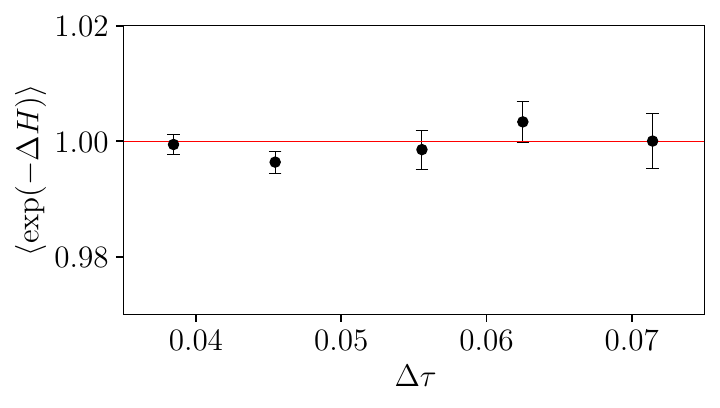}
     \end{subfigure}
     \hfill
     \begin{subfigure}[b]{0.45\textwidth}
         \centering
         \includegraphics[width=0.9\textwidth]{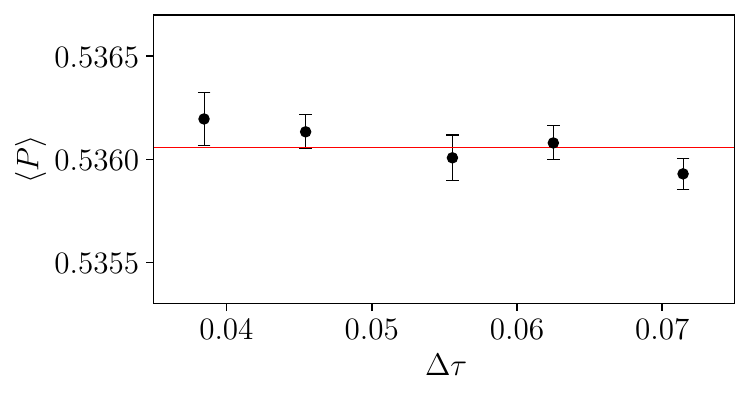}
     \end{subfigure}
        \caption{Test of Creutz equality~\cite{Creutz:1988wv} (left panel), $\langle\exp(-\Delta H)\rangle = 1$ dependence of $\langle\exp(-\Delta H)\rangle$ on the time–step, $\Delta \tau$, in the MD integration, for $N=2$, $N_{\rm f}=0$, and $N_{\rm as}=4$. Test of independence of the plaquette (right panel) on the time–step $\Delta \tau$.  The relevant parameters are the trajectory length $\tau = 1$, number of steps $n_{\rm steps} = 14, 16, 18, 22, 26$ ($\Delta \tau=\tau/n_{\rm steps}$), for an ensemble with lattice volume $\tilde V/a^4=8^4$, 
        $\beta = 6.8$, and $am_0^{\rm as}  = -0.6$. The horizontal line in the right panel represents to the plaquette value obtained averaging over trajectories having different a number of step values, $n_{\rm steps}$. See, for comparison, Ref.~\cite{DelDebbio:2008zf}. (Figure taken from Ref.~\cite{Bennett:2023gbe}).}
        \label{Fig:Creutzandplaq}
\end{figure}

\begin{figure}
     \centering
     \begin{subfigure}[b]{0.45\textwidth}
         \centering
         \includegraphics[width=0.9\textwidth]{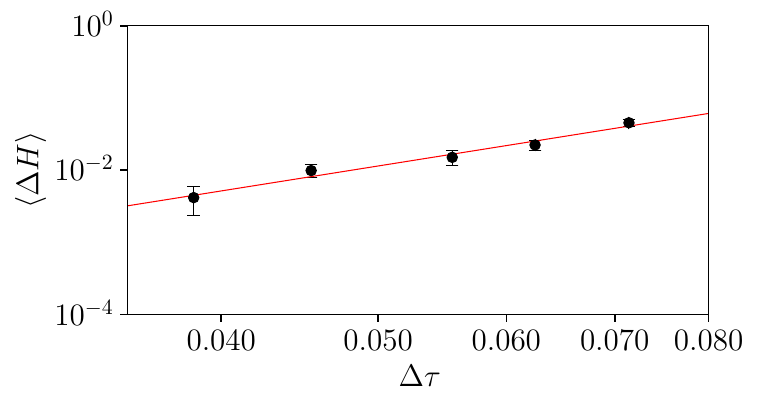}
     \end{subfigure}
     \hfill
     \begin{subfigure}[b]{0.45\textwidth}
         \centering
         \includegraphics[width=0.9\textwidth]{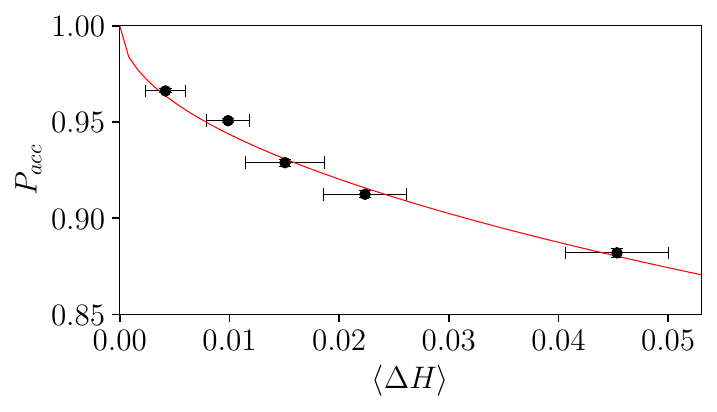}
     \end{subfigure}
        \caption{Left panel: Dependence of $\langle \Delta H \rangle$ on the time-step,  $\Delta \tau$, used for the MD integration. Right panel: Test of the relation between acceptance probability and $\Delta H$~\cite{Gupta:1990ka}.  The theory considered and the relevant parameters of this study are the same as Fig.~\ref{Fig:Creutzandplaq}. These tests follow closely Ref.~\cite{DelDebbio:2008zf}. (Figure taken from Ref.~\cite{Bennett:2023gbe}).}
        \label{Fig:DHandacc}
\end{figure}


\section{Tests of the algorithm}
\label{sec:preliminaryresults}
Focussing primarily on the \spfour theory with $N_{\rm f}=0$, and $N_{\rm as}=4$ Dirac fermions, we perform preliminary algorithm tests, to check the correct implementation of the new code. First of all, we check the sanity of the integrators we use for the molecular dynamics (MD) evolution. The numerical results are presented in Figs.~\ref{Fig:Creutzandplaq} and ~\ref{Fig:DHandacc}. Their correspondent ensemble is obtained evolving the system for $3400$ trajectories and has Madras-Sokal \cite{Madras:1988ei} integrated auto-correlation time $\tau_c = 6.1(2)$.
The first test verifies whether the Creutz equality \cite{Creutz:1988wv} is satisfied. This can be done by measuring the value of the Hamiltonian, $\Delta H$, before and after each trajectory in the HMC evolution to find $\left\langle \exp\left(-\Delta H\right)\right\rangle=1$. This is supported by our numerical results (left panel of Fig.~\ref{Fig:Creutzandplaq}). As a second test, we verify that quantities computed through hybrid Monte-Carlo (HMC and RHMC) updates do not depend on the MD time-size step, $\Delta \tau$, as our updates are obtained through exact algorithms. To do so, we use the elementary plaquette and verify (right panel of Fig.~\ref{Fig:Creutzandplaq}) the independence of such quantity on $\Delta \tau$. As a third test, we verify the relation between $\langle \Delta H\rangle$ and $\Delta \tau$: for a second-order integrator it is supposed to scale as $\langle \Delta H\rangle \propto (\Delta \tau)^4$. In left panel of Fig.~\ref{Fig:DHandacc}, we show the lattice result, together with a best-fit to the curve $\log \langle \Delta H \rangle = {\cal K}_1 \,\log (\Delta \tau) + {\cal K}_2$, with ${\cal K}_1 = 3.6(4)$. The fit value of the reduced is $\chi^2/N_{\rm d.o.f.}=0.6$, and ${\cal K}_1$ is compatible with $4$. As a last check, we show in right panel of Fig.~\ref{Fig:DHandacc} results confirming the predicted relation for the 
acceptance probability, $P_{\rm acc}={\rm erfc}\left(\frac{1}{2}\sqrt{\left\langle\Delta\frac{}{} H\right\rangle}\right)$ ~\cite{Gupta:1990ka}.

\begin{figure}
\includegraphics[width=0.55\textwidth]{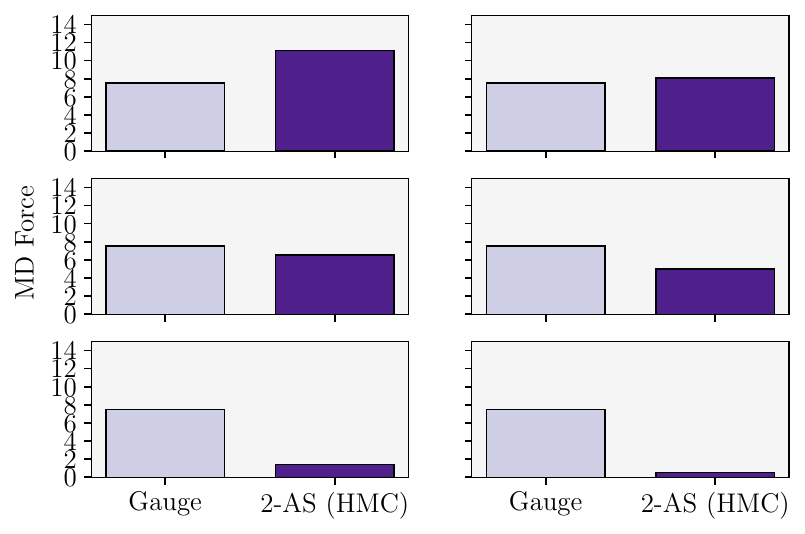}
\centering
\caption{Field contribution to the MD force for the theory with $N=2$, $N_{\rm f}=0$, and $N_{\rm as}=4$, on isotropic lattice with $\tilde V=(8a)^4$, and lattice coupling  $\beta = 6.8$.
The two blocks are respectively indicating the gauge (light shading, left) and the fermion (dark shading, right) contribution. Fermion contributions are summed over flavor. The six panels correspond to different choices of bare mass: $am_0^{\rm as} = -0.9,\, -0.1,\,+0.6,\,+1.8,\,+15,\,+50$ (left to right, top to bottom). The results are normalised so that the gauge contribution is held constant. The computation of the force contributions is made in Ref.~\cite{DelDebbio:2008zf}. (Figure taken from Ref.~\cite{Bennett:2023gbe}). }
\label{Fig:MD}       
\end{figure}

We monitored the contribution to the MD of the fields, and how it changes with bare fermion masses. We show in Fig.~\ref{Fig:MD} the force split in gauge and fermion contributions, $F(x,\mu) = F_g(x,\mu) + F_f(x,\mu)$. The latter includes all four fermions.
As shown in Fig.~\ref{Fig:MD}, for large and positive values of $a m_0^{\rm as}$ the dynamics is led by the gauge degrees of freedom,
as one would expect from a system approaching the quenched regime. Conversely, decreasing the mass, the fermion contribution increases and for negative values of the Wilson bare mass corresponding to small PCAC masses, the fermion contribution dominates.

\begin{figure}
\includegraphics[width=0.7\textwidth]{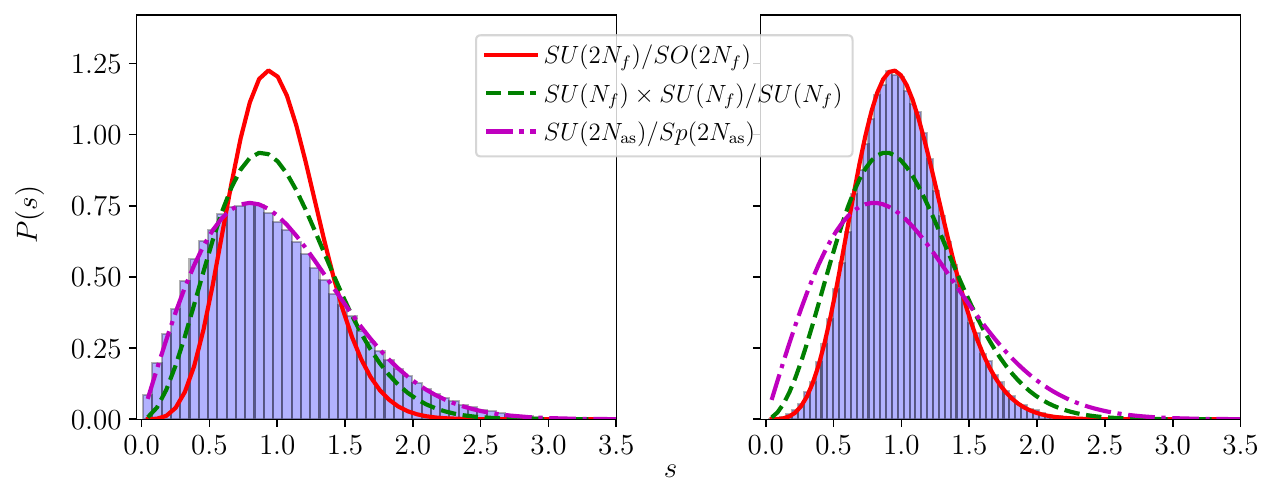}
\centering
\caption{Distribution of the unfolded density of spacing between subsequent eigenvalues of the hermitian Dirac-Wilson operator $Q_m = \gamma_5 D_m$, obtained as in Ref.~\cite{Cossu:2019hse}, and comparison with predictions from chRMT~\cite{Verbaarschot:1994qf}, computed in the quenched approximation, with ensembles having $\beta = 8.0$, $am_0 = -0.2$, and lattice volume $\tilde V=(4a)^4$, in the $Sp(4)$ theory. The left panel shows the case of fermions transforming in the fundamental representation, and the right is for fermions in the 2-index antisymmetric one. The numbers of configurations are $N_{\rm{conf, f}} = 88$ and $N_{\rm{conf, as}} = 47$, while the number of eigenvalues in each configuration
used is $3696$ for fundamental fermions and $5120$ for antisymmetric fermions. (Figure taken from Ref.~\cite{Bennett:2023gbe}).}
\label{Fig:HDW}       
\end{figure}

\begin{figure}
     \centering
     \begin{subfigure}[b]{0.49\textwidth}
         \centering
         \includegraphics[width=0.85\textwidth]{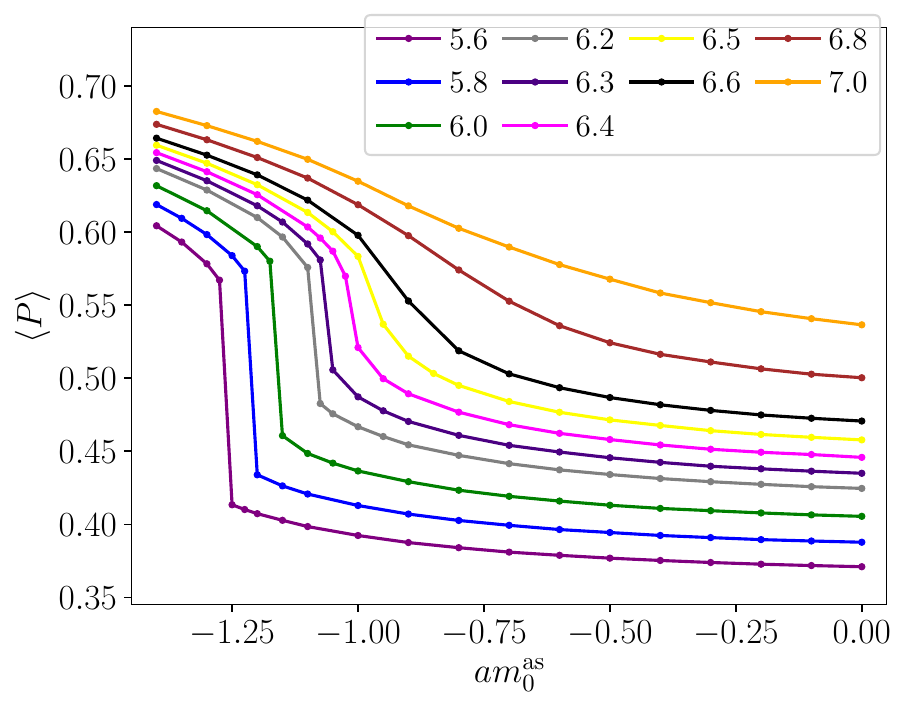}
     \end{subfigure}
     \hfill
     \begin{subfigure}[b]{0.49\textwidth}
         \centering
         \includegraphics[width=0.85\textwidth]{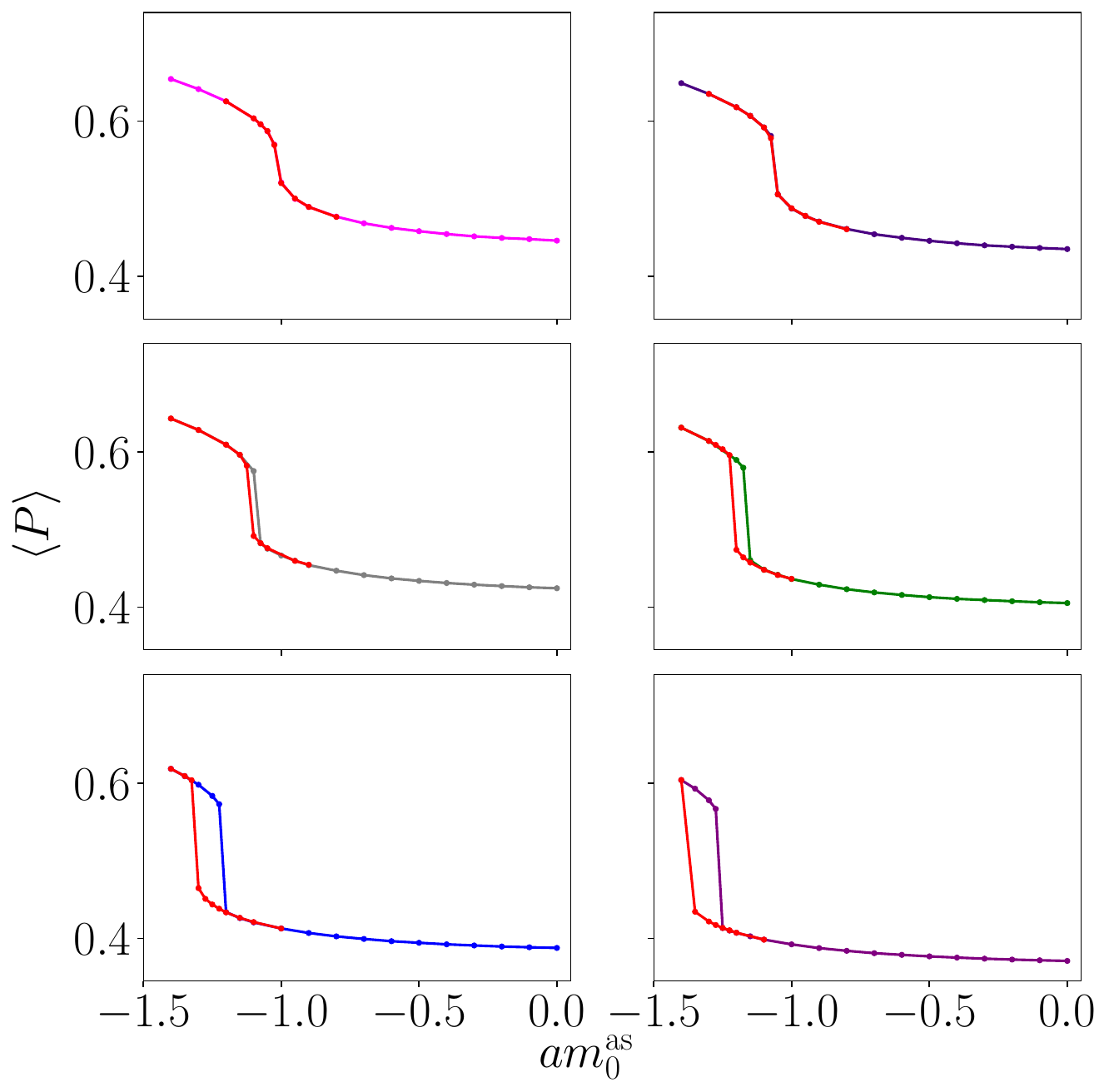}
     \end{subfigure}
        \caption{Left panel: parameter scan of the $Sp(4)$ theory with $N_{\rm f}=0$, $N_{\rm as}=4$ fermions, with ensembles generated from a cold start, using the HMC. We show the value of
  the average plaquette, $\langle P \rangle$, as a function of the bare mass, for a few representative values of the coupling $\beta$ as shown in the legend.  The lattice size is $\tilde V=(8a)^4$, and each point is obtained by varying the lattice coupling  $\beta = 5.6 \hbox{ to } 7.0$. Right panel: Hysteresis between hot (red) and cold (other colors) starts for the $Sp(4)$ theory. The lattice coupling is $\beta = 6.4 \hbox{ to } 5.6$ (left to right, and top to bottom). (Figure taken from Ref.~\cite{Bennett:2023gbe}).}
        \label{Fig:Sp4scan}
\end{figure}

As a last test, we verify that our implementation of the Wilson-Dirac operators is correct. We consider the \spfour theory with quenched fermions in either the fundamental or 2-index antisymmetric representation. We compute the spectrum of eigenvalues of the hermitian Wilson-Dirac operator $Q_m = \gamma_5 D_m$. Following the procedure discussed in Ref.~\cite{Cossu:2019hse}, we compute the distribution of the unfolded density of spacing, $P(s)$, and compare our results to the predictions of chiral Random Matrix Theory~\cite{Verbaarschot:1994qf}. 
The unfolded density of spacing is
\begin{equation}
\label{eq:folded_density}
P(s) = N_{\tilde{\beta}} s^{\tilde{\beta}} \exp\left(-c_{\tilde{\beta}} s^2\right)\,,  \quad  \text{where} \quad N_{\tilde{\beta}} = 2 \dfrac{\Gamma^{\tilde{\beta}+1}\left( \frac{\tilde{\beta}}{2} + 1  \right)}{\Gamma^{\tilde{\beta}+2}\left( \frac{\tilde{\beta}+1}{2}  \right)} , \, c_{\tilde{\beta}} = \dfrac{\Gamma^{2}\left( \frac{\tilde{\beta}}{2} + 1  \right)}{\Gamma^{2}\left( \frac{\tilde{\beta}+1}{2}  \right)}\,,
\end{equation}
where $\tilde{\beta}$ is the Dyson index. As the spectrum is linked to the chiral symmetry-breaking pattern, the distribution $P(s)$ discriminates between the symmetry-breaking patterns associated to different representations of groups. Due to this property, the Dyson index takes different values: $\tilde{\beta} = 4$ corresponds to $SU(2N_{f}) \to SO(2N_{f})$, $\tilde{\beta} = 2$ to $SU(N_f) \times SU(N_f) \to SU(N_f)$,  and $\tilde{\beta} = 1$ to $SU(2N_{f}) \to Sp(2N_{f})$. To compare our results on the lattice with Eq.~(\ref{eq:folded_density}), we compute the eigenvalues of $Q_m$ for $N_{\rm conf}$ configurations. Then, following the procedure described in Ref.~\cite{Cossu:2019hse}, we find the \textit{discretised} unfolded density of spacings, $P(s)$.
In Fig.~\ref{Fig:HDW}, we show our numerical results: one finds a distribution that is compatible with the expected symmetry-breaking patterns. The observed agreement with the predicted model gives us strong indication that we correctly implemented the Wilson-Dirac operators. \footnote{The correctness of the Wilson–Dirac operator could also be checked via consistency with the Feynman rules, both by using the free propagator obtained from Fourier transforms, and by comparing the results of gauge transformation. This has not been done in this work.}
\\We performed a lattice parameter space scan, to identify possible phase transitions happening while varying lattice parameters, by studying the average plaquette, $\langle P \rangle$, and its hysteresis. Performing such a study we know where there is no bulk phase transition, and one can safely perform lattice numerical calculations. The left panel of Fig.~\ref{Fig:Sp4scan} displays the average plaquette, $\langle P \rangle$, in ensembles generated using a cold start. For this theory, the average plaquette is a smooth function everywhere, except for precise values of $\beta_*$ and  $a m_0^{{\rm as}\,\ast}$, where it shows an abrupt change---this gives us a strong indication of a first-order, bulk phase transition. 
As a further verification, we re-generated the same ensembles with hot starts and repeated the measurements using the same lattice parameters. The right panel of figure~\ref{Fig:Sp4scan} shows the comparison of the average plaquette values, $\langle P \rangle$, between hot and cold starts, using the same bare lattice parameters and hysteresis is clearly visible, for $\beta< \beta_* \simeq 6.4$. 

We adopt a scale setting procedure that makes use of the Wilson flow.~\cite{Luscher:2013vga}. One introduces the fifth-dimension, flow time, $t$, and solves the defining diffusion differential equation $\dfrac{\mathrm{d} B_\mu(x,\,t)}{\mathrm{d}t} = D_\nu G_{\nu\mu}(x,\,t)$,
with the boundary conditions, $B_\mu(x,\,0)=U_\mu(x)$. 
Then, one defines the quantities ${\cal E}(t) \equiv \frac{t^2}{2} \left\langle \mathrm{Tr} \,
    \left[ G_{\mu\nu}(t) G_{\mu\nu}(t)\right]\right\rangle\,, {\cal W}(t)  \equiv  t \frac{d}{dt}  \mathcal{E}(t)$
and introduces a prescription $\left. \mathcal{E}(t)\right|_{t=t_0} = \mathcal{E}_0$ that sets the scale $t_0$, or alteratively $\left. {\cal W}(t)\right|_{t=w_0^2} = \mathcal{W}_0$ that sets the scale $w_0^2$. Both $\mathcal{E}_0$ and $\mathcal{W}_0$ are chosen conventionally.
\begin{figure}
     \centering
     \begin{subfigure}[b]{0.49\textwidth}
         \centering
         \includegraphics[width=0.8\textwidth]{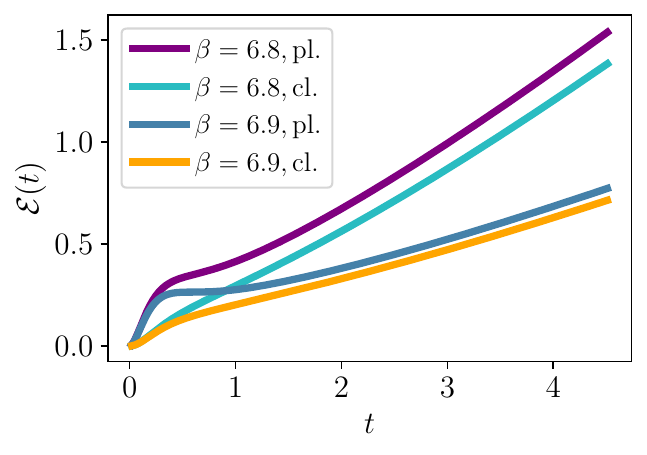}
     \end{subfigure}
     \hfill
     \begin{subfigure}[b]{0.49\textwidth}
         \centering
         \includegraphics[width=0.8\textwidth]{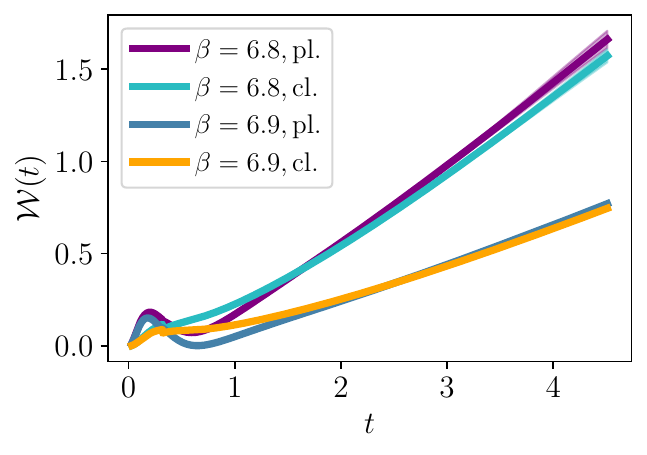}
     \end{subfigure}
        \caption{Wilson Flow~\cite{Luscher:2013vga} energy density ${\cal E}(t)$ (left panel) and ${\cal W}(t)$ (right), computed as in Refs.~\cite{Bennett:2017kga,Bennett:2022ftz}, from the standard (pl) and the clover-leaf (cl) plaquette, for the $Sp(4)$ theory with $N_{\rm as}=4$ fermions transforming in the 2-index antisymmetric representation. The lattice size is $\tilde V = (12a)^4$,  $\beta = 6.8$ and  $6.9$, $a m_0^{\rm as} = -0.8$. The time step is $0.01$, $t_{max} = 4.5$ to reduce finite-size effects. We choose ${\cal W}_0 = \frac{1}{2} C_2(F)$ for the scale setting. The corresponding values of $w_0$ from the plaquette and the clover-leaf are $w_{0, pl.} = 1.485(3)$ and $w_{0, cl.} = 1.495(2)$ for $\beta = 6.8$ and $w_{0, pl.} = 2.005(2)$ and $w_{0, cl.} = 2.026(2)$ for $\beta = 6.9$. We have set $a=1$, for notational convenience. (Figure taken from Ref.~\cite{Bennett:2023gbe}).}
        \label{Fig:WF}
\end{figure}
In Fig.~\ref{Fig:WF} we show ${\cal E}(t)$ and ${\cal W}(t)$ as functions of the flow time, $t$. On the lattice, the calculation of ${\cal E}(t)$ and ${\cal W}(t)$ depends on a definition of $G_{\mu\nu}$, and we display explicitly two choices: the elementary, $\mathcal{P}_{\mu\nu}(x)$, and the clover-leaf plaquette, $C_{\mu\nu}(x)$.
The plots show results agreeing with previous findings in the literature, according to which at early flow times ${\cal E}(t)$ and ${\cal W}(t)$ strongly differ due to UV fluctuations. Then, the cut-off effects are smoothened and the two curves become closer to each other. Moreover, we notice that the function ${\cal W}(t)$ displays a milder dependence. For this reason, we set the scale $w_0$ using ${\cal W}(t)$, by 
conventionally setting ${\cal W}_0=\frac{1}{2}C_2(F)$.  Having set the scale, one can define the topological charge. For gauge configurations generated by Monte Carlo simulation, this observable is dominated by UV fluctuations, hence it will be regulated defining it through $B_\mu(x,t)$, obtaining $Q_L(t)\equiv\frac{1}{32 \pi^2} \varepsilon^{\mu\nu\rho\sigma}\sum_x\hbox{Tr}\,\left[{\cal C}_{\mu\nu}(x,t){\cal C}_{\rho\sigma}(x,t)\right]$.
\begin{figure}
     \centering
     \begin{subfigure}[b]{0.49\textwidth}
         \centering
         \includegraphics[width=0.9\textwidth]{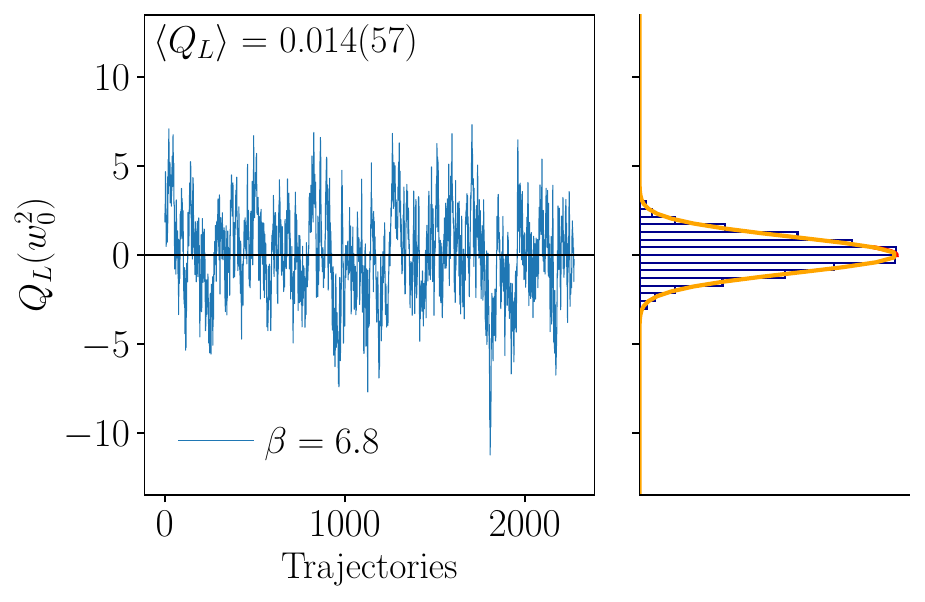}
     \end{subfigure}
     \hfill
     \begin{subfigure}[b]{0.49\textwidth}
         \centering
         \includegraphics[width=0.9\textwidth]{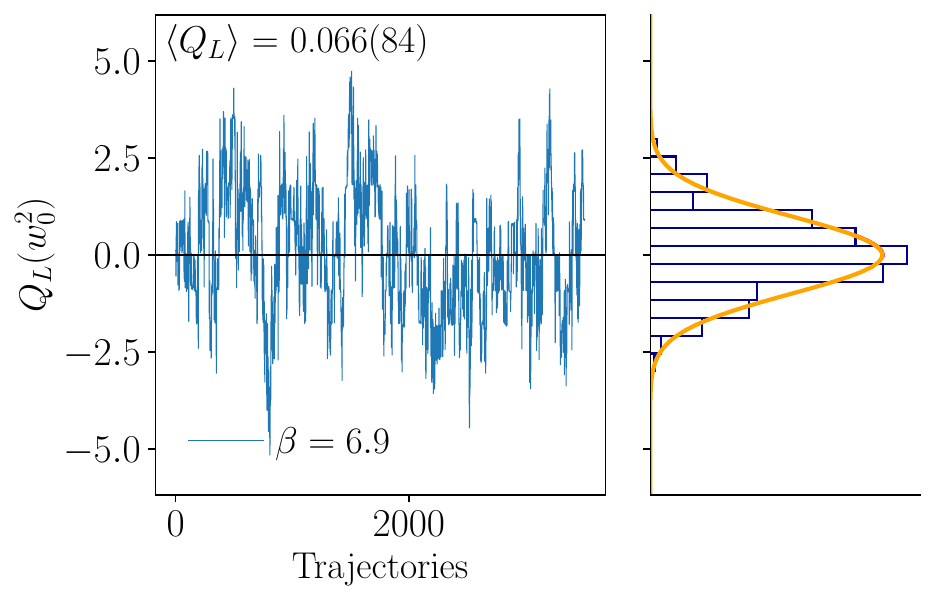}
     \end{subfigure}
        \caption{Evolution with the ensemble trajectories 
   of the  topological charge computed at flow time $t = w_0^2$, as shown in Ref.~\cite{Bennett:2022ftz}, for the $Sp(4)$ theory with $N_{\rm as}=4$ fermions transforming in the 2-index antisymmetric representation. The lattice parameters and size are the same as Fig.~\ref{Fig:WF}. The histograms of the measurements (right panels) are compatible with a  normal distribution centered at zero, with reduced chi-square $\chi^2 / N_{\rm d.o.f }= \tilde{\chi}^2 = 1.1$ for both panels. The integrated autocorrelation time computed using the Madras-Sokal windowing algorithm is $\tau_Q = 31(3)$ (left panel)
   and $\tau_Q = 238(12)$ (right panel). (Figure taken from Ref.~\cite{Bennett:2023gbe}).}
        \label{Fig:Q}
\end{figure}
In Fig.~\ref{Fig:Q} we display the value of $Q_L(t=w_0^2)$,
for the same ensembles of the Wilson flow study. There is no evidence of topological freezing, as the Madras-Sokal integrated autocorrelation time~\cite{Madras:1988ei}, $\tau_Q$, of the topological charge turns out to be many orders of magnitude smaller than the number of trajectories.


\section{Summary and outlook}
\label{sec:conclusions}
Symplectic gauge theories have a variety of phenomenological applications in many contexts such as Composite Higgs Models and top partial compositeness~\cite{Barnard:2013zea,Ferretti:2013kya}, strongly interacting dark matter models~\cite{PhysRevLett.113.171301, PhysRevLett.115.021301}. We developed and tested new software, embedded into the Grid environment to take full advantage of its flexibility. We reported the positive results of our tests of the algorithms, particularly focusing on the $Sp(4)$ theory coupled to $N_{as}= 4$ (Dirac) fermions transforming in the antisymmetric representation. This work and the software we developed for it set the stage needed to explore and quantify future large-scale studies---among them, the study of the conformal window extent in strongly coupled gauge theories with matter field content. Moreover, the tools we developed will be useful in the context of the recent literature discussing the spectroscopy of \spn theories with various representations (see, e.g. Refs.\cite{Bennett:2017kga,Bennett:2019jzz,Bennett:2022yfa}), and can be further extended by applying new techniques based on the spectral densities~\cite{Hansen:2019idp}.

\acknowledgments
The work of EB, JL and BL has been funded by the ExaTEPP project EP/X017168/1. The work of EB and JL has also been supported by the UKRI Science and Technology Facilities Council (STFC) Research Software Engineering Fellowship EP/V052489/1. The work of NF has been supported by the STFC Consolidated Grant No. ST/X508834/1. The work of PB was supported in part by US DOE Contract DESC0012704(BNL), and in part by the Scientific Discovery through Advanced Computing (SciDAC) program LAB 22-2580. The work of DKH was supported by Basic Science Research Program through the National Research Foundation of Korea (NRF) funded by the Ministry of Education (NRF-2017R1D1A1B06033701). The work of LDD and AL was supported by the ExaTEPP project EP/X01696X/1. The work of JWL was supported in part by the National Research Foundation of Korea (NRF) grant funded by the Korea government(MSIT) (NRF-2018R1C1B3001379) and by IBS under the project code, IBS-R018-D1. The work of DKH and JWL was further supported by the National Research Foundation of Korea (NRF) grant funded by the Korea government (MSIT) (2021R1A4A5031460). The work of CJDL is supported by the Taiwanese NSTC grant 109-2112-M-009-006-MY3. DV is supported by a STFC new applicant scheme grant. The work of BL and MP has been supported in part by the STFC Consolidated Grants No. ST/P00055X/1, ST/T000813/1, and ST/X000648/1. BL, MP, AL and LDD received funding from the European Research Council (ERC) under the European Union’s Horizon 2020 research and innovation program under Grant Agreement No. 813942. The work of BL is further supported in part by the EPSRC ExCALIBUR programme ExaTEPP (project EP/X017168/1), by the Royal Society Wolfson Research Merit Award WM170010 and by the Leverhulme Trust Research Fellowship No. RF-2020-4619. LDD is supported by the UK Science and Technology Facility Council (STFC) grant ST/P000630/1. Numerical simulations have been performed on the Swansea SUNBIRD cluster (part of the Supercomputing Wales project) and AccelerateAI A100 GPU system, and on the DiRAC Extreme Scaling service at the University of Edinburgh. Supercomputing Wales and AccelerateAI are part funded by the European Regional Development Fund (ERDF) via Welsh Government. The DiRAC Extreme Scaling service is operated by the Edinburgh Parallel Computing Centre on behalf of the STFC DiRAC HPC Facility (www.dirac.ac.uk). This equipment was funded by BEIS capital funding via STFC capital grant ST/R00238X/1 and STFC DiRAC Operations grant ST/R001006/1. DiRAC is part of the National e-Infrastructure. \\
{\bf Research Data Access Statement and Open Access Statement}---The data shown in this manuscript can be downloaded from the data release of Ref.~\cite{Bennett:2023gbe}. For the purpose of open access, the authors have applied a Creative Commons 
Attribution (CC BY) licence to any Author Accepted Manuscript version arising.


\bibliographystyle{JHEP}
\bibliography{references.bib}

\providecommand{\href}[2]{#2}\begingroup\raggedright\begin{thebibliography}{10}

\bibitem{Bennett:2017kga}
{E. Bennett et al.}, \emph{{Sp(4) gauge theory on the lattice: towards SU(4)/Sp(4) composite Higgs (and beyond)}}, \href{https://doi.org/10.1007/JHEP03(2018)185}{\emph{JHEP} {\bfseries 03} (2018) 185} [\href{https://arxiv.org/abs/1712.04220}{{\ttfamily 1712.04220}}].

\bibitem{Bennett:2019jzz}
{E. Bennett et al.}, \emph{{Sp(4) gauge theories on the lattice: $N_f=2$ dynamical fundamental fermions}}, \href{https://doi.org/10.1007/JHEP12(2019)053}{\emph{JHEP} {\bfseries 12} (2019) 053} [\href{https://arxiv.org/abs/1909.12662}{{\ttfamily 1909.12662}}].

\bibitem{Bennett:2019cxd}
{E. Bennett et al.}, \emph{{$Sp(4)$ gauge theories on the lattice: quenched fundamental and antisymmetric fermions}}, \href{https://doi.org/10.1103/PhysRevD.101.074516}{\emph{Phys. Rev. D} {\bfseries 101} (2020) 074516} [\href{https://arxiv.org/abs/1912.06505}{{\ttfamily 1912.06505}}].

\bibitem{Bennett:2022yfa}
{E. Bennett et al.}, \emph{{Lattice studies of the Sp(4) gauge theory with two fundamental and three antisymmetric Dirac fermions}}, \href{https://doi.org/10.1103/PhysRevD.106.014501}{\emph{Phys. Rev. D} {\bfseries 106} (2022) 014501} [\href{https://arxiv.org/abs/2202.05516}{{\ttfamily 2202.05516}}].

\bibitem{Bennett:2023wjw}
{E. Bennett et al.}, \emph{{Sp(2N) Lattice Gauge Theories and Extensions of the Standard Model of Particle Physics}}, \href{https://doi.org/10.3390/universe9050236}{\emph{Universe} {\bfseries 9} (2023) 236} [\href{https://arxiv.org/abs/2304.01070}{{\ttfamily 2304.01070}}].

\bibitem{Bennett:2023rsl}
{E. Bennett et al.}, \emph{{Singlets in gauge theories with fundamental matter}},  \href{https://arxiv.org/abs/2304.07191}{{\ttfamily 2304.07191}}.

\bibitem{Barnard:2013zea}
{J. Barnard et al.}, \emph{{UV descriptions of composite Higgs models without elementary scalars}}, \href{https://doi.org/10.1007/JHEP02(2014)002}{\emph{JHEP} {\bfseries 02} (2014) 002} [\href{https://arxiv.org/abs/1311.6562}{{\ttfamily 1311.6562}}].

\bibitem{Ferretti:2013kya}
{G. Ferretti et al.}, \emph{{Fermionic UV completions of Composite Higgs models}}, \href{https://doi.org/10.1007/JHEP03(2014)077}{\emph{JHEP} {\bfseries 03} (2014) 077} [\href{https://arxiv.org/abs/1312.5330}{{\ttfamily 1312.5330}}].

\bibitem{Panico:2015jxa}
{G. Panico et al.}, \emph{{The Composite Nambu-Goldstone Higgs}}, vol.~913, Springer (2016), \href{https://doi.org/10.1007/978-3-319-22617-0}{10.1007/978-3-319-22617-0}, [\href{https://arxiv.org/abs/1506.01961}{{\ttfamily 1506.01961}}].

\bibitem{Cacciapaglia:2019bqz}
{Cacciapaglia et al.}, \emph{{Light scalars in composite Higgs models}}, \href{https://doi.org/10.3389/fphy.2019.00022}{\emph{Front. in Phys.} {\bfseries 7} (2019) 22}.

\bibitem{Kaplan:1991dc}
D.B.~Kaplan, \emph{{Flavor at SSC energies: A New mechanism for dynamically generated fermion masses}}, \href{https://doi.org/10.1016/S0550-3213(05)80021-5}{\emph{Nucl. Phys. B} {\bfseries 365} (1991) 259}.

\bibitem{Boyle:2015tjk}
{P. Boyle et al.}, \emph{{Grid: A next generation data parallel C++ QCD library}},  \href{https://arxiv.org/abs/1512.03487}{{\ttfamily 1512.03487}}.

\bibitem{Boyle:2016lbp}
{P. Boyle et al.}, \emph{{Grid: A next generation data parallel C++ QCD library}}, \href{https://doi.org/10.22323/1.251.0023}{\emph{PoS} {\bfseries LATTICE2015} (2016) 023}.

\bibitem{sp2ngrid}
{Grid contributors}, \emph{{GRID, Data parallel C++ mathematical object library -- Sp2n version}}, {\emph{doi:10.5281/zenodo.8136357} }.

\bibitem{Creutz:1988wv}
M.~Creutz, \emph{{Global Monte Carlo algorithms for many-fermion systems}}, \href{https://doi.org/10.1103/PhysRevD.38.1228}{\emph{Phys. Rev. D} {\bfseries 38} (1988) 1228}.

\bibitem{DelDebbio:2008zf}
{Del Debbio et al.}, \emph{{Higher representations on the lattice: Numerical simulations. SU(2) with adjoint fermions}}, \href{https://doi.org/10.1103/PhysRevD.81.094503}{\emph{Phys. Rev. D} {\bfseries 81} (2010) 094503} [\href{https://arxiv.org/abs/0805.2058}{{\ttfamily 0805.2058}}].

\bibitem{Bennett:2023gbe}
E.~Bennett et~al., \emph{{Symplectic lattice gauge theories on Grid: approaching the conformal window}},  \href{https://arxiv.org/abs/2306.11649}{{\ttfamily 2306.11649}}.

\bibitem{Gupta:1990ka}
{Gupta et al.}, \emph{{The Acceptance Probability in the Hybrid Monte Carlo Method}}, \href{https://doi.org/10.1016/0370-2693(90)91790-I}{\emph{Phys. Lett. B} {\bfseries 242} (1990) 437}.

\bibitem{Madras:1988ei}
N.~Madras and A.D.~Sokal, \emph{{The Pivot algorithm: a highly efficient Monte Carlo method for selfavoiding walk}}, \href{https://doi.org/10.1007/BF01022990}{\emph{J. Statist. Phys.} {\bfseries 50} (1988) 109}.

\bibitem{Cossu:2019hse}
{G. Cossu et al.}, \emph{{Strong dynamics with matter in multiple representations: $\mathrm {SU}(4)$ gauge theory with fundamental and sextet fermions}}, \href{https://doi.org/10.1140/epjc/s10052-019-7137-1}{\emph{Eur. Phys. J. C} {\bfseries 79} (2019) 638} [\href{https://arxiv.org/abs/1904.08885}{{\ttfamily 1904.08885}}].

\bibitem{Verbaarschot:1994qf}
J.J.M.~Verbaarschot, \emph{{The Spectrum of the QCD Dirac operator and chiral random matrix theory: The Threefold way}}, \href{https://doi.org/10.1103/PhysRevLett.72.2531}{\emph{Phys. Rev. Lett.} {\bfseries 72} (1994) 2531} [\href{https://arxiv.org/abs/hep-th/9401059}{{\ttfamily hep-th/9401059}}].

\bibitem{Luscher:2013vga}
M.~L\"uscher, \emph{{Future applications of the Yang-Mills gradient flow in lattice QCD}}, \href{https://doi.org/10.22323/1.187.0016}{\emph{PoS} {\bfseries LATTICE2013} (2014) 016} [\href{https://arxiv.org/abs/1308.5598}{{\ttfamily 1308.5598}}].

\bibitem{Bennett:2022ftz}
{E. Bennett et al.}, \emph{{Sp(2N) Yang-Mills theories on the lattice: Scale setting and topology}}, \href{https://doi.org/10.1103/PhysRevD.106.094503}{\emph{Phys. Rev. D} {\bfseries 106} (2022) 094503} [\href{https://arxiv.org/abs/2205.09364}{{\ttfamily 2205.09364}}].

\bibitem{PhysRevLett.113.171301}
Y.~Hochberg and collaborators, \emph{Mechanism for thermal relic dark matter of strongly interacting massive particles}, \href{https://doi.org/10.1103/PhysRevLett.113.171301}{\emph{Phys. Rev. Lett.} {\bfseries 113} (2014) 171301}.

\bibitem{PhysRevLett.115.021301}
Y.~Hochberg and collaborators, \emph{Model for thermal relic dark matter of strongly interacting massive particles}, \href{https://doi.org/10.1103/PhysRevLett.115.021301}{\emph{Phys. Rev. Lett.} {\bfseries 115} (2015) 021301}.

\bibitem{Hansen:2019idp}
{M. Hansen et al.}, \emph{{Extraction of spectral densities from lattice correlators}}, \href{https://doi.org/10.1103/PhysRevD.99.094508}{\emph{Phys. Rev. D} {\bfseries 99} (2019) 094508} [\href{https://arxiv.org/abs/1903.06476}{{\ttfamily 1903.06476}}].

\end{thebibliography}\endgroup

\end{document}